\documentstyle[11pt,newpasp,twoside,epsf]{article}
\begin{document}

\title{A Joint model for radio and gamma-ray emission from pulsars}

\author{Qiao,G.J., Lee,K.J.,
    Wang,H.G., and Xu,R.X.}

\affil{Astronomy Department, Peking University,
       Beijing 100871, P.R.China. }

\begin{abstract}
Although pulsars can radiate electromagnetic wave from radio to
gamma ray bands, we still have no a united model to understand the
multi-band emission. In this paper the effort for a joint model is
presented. The inverse Compton scattering (ICS) and a second
acceleration process near the null charge surface are involved to
account for the radio and the gama-ray emission, respectively.
Various kind of pulse profiles and other observational properties
can be reproduced.
\end{abstract}


\section{Introduction}

Pulsar is able to radiate multi-band emission. A wealth of
observational data on radio pulsars has been collected since the
discovery of pulsar. In high energy, 7 gamma-ray pulsars have been
discovered by far. Gamma-ray photons from young pulsars allow the
deepest insight into the properties of high-energy particles and
their interaction with the magnetic fields and photons in pulsar
magnetosphere. To understand the observational facts, polar cap
models are proposed to account for the radio emission and both the
polar cap and outer gap models are developed to explain the
gamma-ray emission. In the past decades these two domains have
been investigated separately. However, both radio and gamma-ray
emissions can be radiated from the same pulsar, such as the Crab
and the Vela pulsars, therefore it is highly necessary to
establish a united model for radio and gamma-ray emissions.

\section{Radio emission from pulsars}

There are two kinds of polar cap models for pulsar radio emission.
One is the inner gap model, which suggests that a gap-type
accelerator could be formed on the polar cap surface (Ruderman \&
Sutherland 1975), the other is space-charge-limited flow, which
suggests that either the negative or the positive ions could flow
freely from the stellar surface (Arons 1983). Although the binding
energy of positive ions is not so high to form an inner gap on
very hot neutron star surface, recent investigations found that
inner gap may still exist in the situation of bare strange star
(Xu et al. 2001) or in some cases of neutron star (Gil et al.
2002).

Any radio model is required to be able to explain the main
observational facts, e.g., the pulse profile, the polarization,
the spectrum and so on. We have proposed an inverse Compton
scattering (ICS) model for radio emission based on the inner gap
scenario (see Qiao \& Lin 1998, hereafter paper I, Xu et al. 2000,
Qiao et al. 2001). Under this model, some important observational
properties can be reproduced: (1) the central (or "core") emission
beam and the conal beams; (2) the location for each emission
component; (3) the linear and circular polarization of individual
and integrated pulses; (4) the pulse profiles changing with the
frequency. The ICS model is involved in our joint model to account
for the radio emission.

\section{Gamma-ray emission from pulsars}

Even if the local charge density $\rho$ of secondary particles
just out of the inner gap is the same as that of local
Goldreich-Julian density $\rho_{gj}$ ($\rho_{gj}=-{\bf \Omega}
\cdot {\bf B }/ (2 \pi c)$), when the particles stream out, the
charge density should be different from the local $\rho_{gj}$,
then the acceleration caused by space-charge-limited flow will
take place. The potential can be written as (Arons 1983)
\begin{equation}
-\nabla ^2 {\phi} =4 \pi (\rho - \rho_{gj})
\end{equation}
In the one-dimensional case, the electric field parallel to the
magnetic field reads (Michel 1974)

\begin{equation}
dE_{\|}/dz =4 \pi (\rho - \rho_{gj})
\end{equation}
In this way the particles will be accelerated effectively near the
null charge surface where $\rho_{gj}=0$ and then radiate
gamma-rays. The geometry of this acceleration location is shown in
Fig.1 schematically. Such an acceleration process is effective
within the bundle of open field lines that intersect the null
charge surface.

The gamma-ray beam and light curve are figured out according to
geometrical relations (Fig.2). Since the relative location of the
secondary acceleration region to the polar cap surface depends on
the inclination angle, the calculated gamma beam can be narrow or
very wide, and the gamma and radio pulses may show various kinds
of phase configuration, from alignment to a wide separation. For
the comparison of calculated results with the observational
profiles readers are referred to Qiao et al. (2002). The details
of physical process are to be presented elsewhere.

\section{Conclusions and discussions}
From the geometrical consideration above one can see that both the
inner and the outer gaps may play significant roles in pulsar
radiation. The key point is that the inner gap acceleration and
the space-charge-limited flow acceleration near the null charge
surface should be taken into account. Low frequency waves supplied
by the inner gap sparking are inverse Compton scattered by the
outgoing relativistic particles to produce the radio emission
(Paper I). The particles then encounter a secondary acceleration
near the null charge surface and emit gamma-rays. Various kinds of
observational properties can be reproduced. Besides some pulsars
can only be observed in gamma-ray, we predict that some pulsars
can not be observed in gamma-ray owing to our line of sight
missing the emission beam. The joint model presented here is
different from the cur-rent polar cap gamma-ray models because the
null charge surface plays the vital role in secondary
acceleration. Our model is also different from the present outer
gap models: the secondary particles are accelerated by the effect
of space-charge-limited flow and the radiation location is
extended inside into the null surface.

\begin{acknowledgements}
We are very grateful to Professor R.N. Manchester for his valuable
discussion.  This work is partly supported by NSF of China, and
the Research Fund for the Doctoral Program Higher Education.
\end{acknowledgements}

\begin {figure}
\plotone {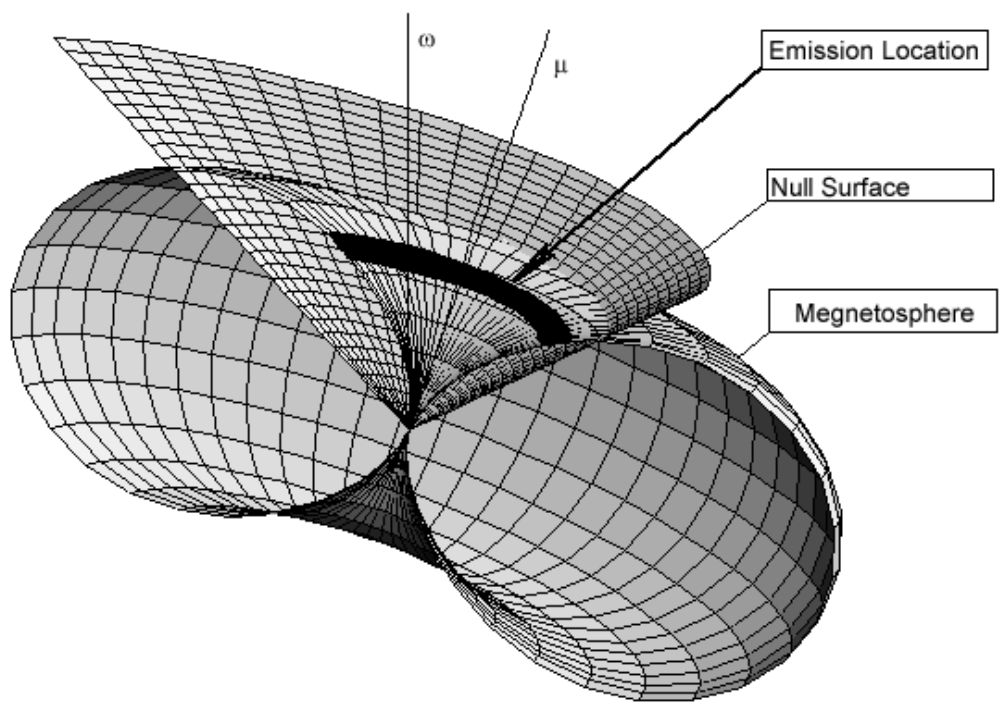} \caption{The null charge surface and the emission
location.}
\end {figure}

\begin {figure}
\plotone {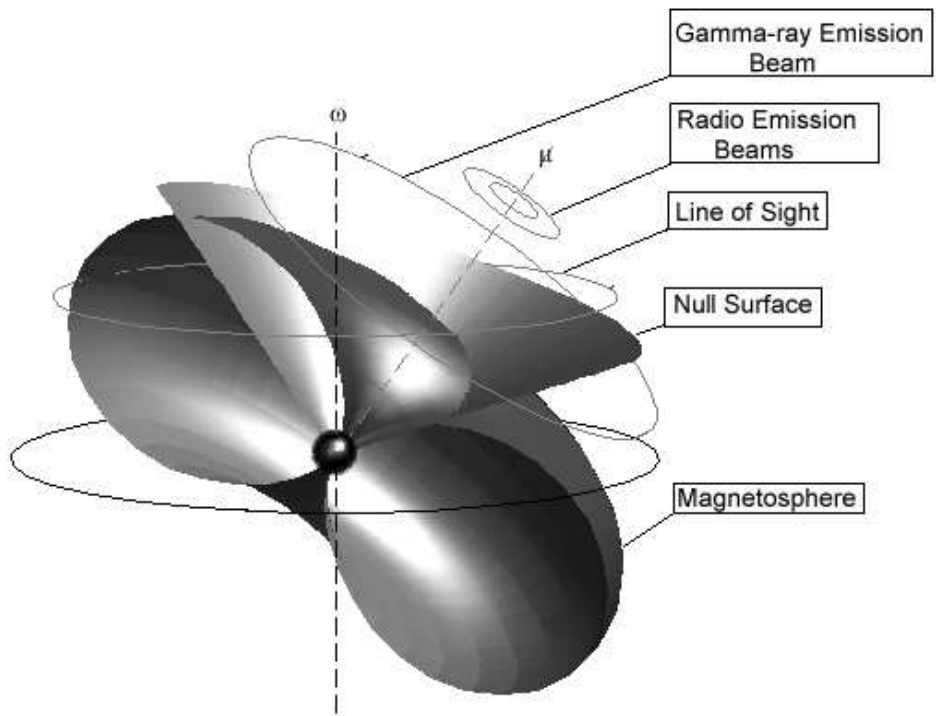} \caption{The pulsar's magnetosphere, null charge
surface and emission beams. It is shown that the gamma-ray beam
can be rather wide with the emission location inside the null
charge surface.}
\end {figure}

\end{document}